\documentclass[aps,prapplied,reprint,superscriptaddress]{revtex4-2}

\usepackage{graphicx}
\usepackage{dcolumn}
\usepackage{bm}
\usepackage{color}
\usepackage{caption}
\usepackage{graphicx, subfig}
\usepackage{epsfig}
\usepackage{ulem}
\usepackage[dvipsnames]{xcolor}
\definecolor{green1}{rgb}{ 0.3882  ,  0.5961  ,  0.3882}
\definecolor{gra}{rgb}{ 0.8  ,  0.8  ,  0.8}
\usepackage[defaultcolor=Maroon]{changes}
\definechangesauthor[name={Tian-Xue Ma},color=BlueGreen]{MTX}
\newdimen\Lmargin
\newdimen\Rmargin
\Rmargin=\paperwidth \advance\Rmargin by-\Lmargin \advance\Rmargin by-\textwidth
\ifdim\Lmargin>\Rmargin \Rmargin=\Lmargin \fi

\usepackage{amssymb}
\usepackage{amsmath}
\usepackage{makecell}
\usepackage[colorlinks=true,linkcolor=blue,citecolor=blue,urlcolor=blue]{hyperref}

\begin{document}


\title{Inverse Design of Topological Interface and Corner States in Two-Dimensional Phononic Crystals via Deep Learning}



\author{Xue-Qian Zhang}
\affiliation{Department of Mechanics, School of Physical Sciences and Engineering,
	Beijing Jiaotong University, Beijing 100044, China}
\affiliation{Instituto Universitario de Matemática Pura y Aplicada,
	Universitat Polit\`ecnica de Val\`encia, 46022, Spain}

\author{Jincheng Tang}
\affiliation{Department of Mechanics, School of Physical Sciences and Engineering,
	Beijing Jiaotong University, Beijing 100044, China}

\author{Tian-Xue Ma}
\email{matx@bjtu.edu.cn}
\affiliation{Department of Mechanics, School of Physical Sciences and Engineering,
	Beijing Jiaotong University, Beijing 100044, China}
\affiliation{Beijing Key Laboratory of Novel Materials Genetic Engineering and Application for Rail Transit, Beijing Jiaotong University, Beijing 100044, China}

\author{Yue-Sheng Wang}
\email{yswang@tju.edu.cn}
\affiliation{Department of Mechanics, School of Physical Sciences and Engineering,
	Beijing Jiaotong University, Beijing 100044, China}
\affiliation{Department of Mechanics, School of Mechanical Engineering,
	Tianjin University, Tianjin 300350, China}

\author{Vin\'icius Fonseca Dal Poggetto}
\email{vfondal@upvnet.upv.es}
\affiliation{Instituto Universitario de Matemática Pura y Aplicada,
	Universitat Polit\`ecnica de Val\`encia, 46022, Spain}

\author{Vicent Romero-Garc\'ia}
\email{virogar1@mat.upv.es}
\affiliation{Instituto Universitario de Matemática Pura y Aplicada,
	Universitat Polit\`ecnica de Val\`encia, 46022, Spain}

\date{\today}

\begin{abstract}
Topological states in phononic crystals (PnCs) exhibit remarkable localization and robustness, making them highly advantageous for applications including waveguiding, energy harvesting, and acoustic sensing. However, the precise tailoring of specific topological states remains a considerable challenge. Prior studies have largely focused on band gap characteristics and their associated topological properties, from which topological states are inferred indirectly. To address this limitation, we introduce a deep learning framework for the direct engineering of topological interface states (TISs) and topological corner states (TCSs) at predetermined frequencies within two-dimensional topological PnCs. The unit cells, represented as 32 $\times$ 32 pixel matrices with free-form geometries, are compressed into a 10-dimensional latent space via a conditional variational autoencoder, thereby improving design efficiency. Forward prediction models employing multilayer perceptrons are developed to estimate band gap boundaries, their topological properties, TIS gaps, and TCS frequencies. For the inverse design, we propose a Latent-Consistent Tandem Network that enables both direct and multi-objective design of target structural properties and topological states. By integrating the filling ratio as an additional design parameter, the framework enables multi-objective inverse design that simultaneously considers physical properties and geometric features. The approach generates PnCs with prescribed topological band gap boundaries and topological state frequencies, achieving correlation coefficients exceeding 0.96 between the target and achieved properties. Furthermore, the framework supports “one-to-many” inverse design and allows modulation of the filling ratio in the generated unit cells. Customized designs of TIS gaps and TCS frequencies are successfully realized and experimentally validated. This work underscores the potential of deep learning techniques for the intelligent design of topological PnCs and establishes a novel paradigm for the development of engineered functional materials.
\end{abstract}


\maketitle

\section{Introduction}

Topological phononic crystals (PnCs) have attracted significant attention for their ability to support robust wave localization and transport~\cite{wang2015topological,laforge2021acoustic,xu2023realization,li2023direct}. By interfacing PnCs with distinct topological properties, topological interface states (TISs) and topological corner states (TCSs) can emerge within band gaps (BGs)~\cite{gao2025simplified}.
In these systems, topologically protected states of acoustic waves have shown great potential for applications in waveguiding \cite{sun2025research, yu2026directional, khelif2004guiding, chen2025broadband,gao2021broadband}, energy harvesting \cite{zhang2026energy, li2023acoustic, tang2023topological}, acoustic sensing \cite{zaky2025coupling, fathy2024design}, filtering \cite{chen2017acoustic, wang2025frequency, lin2025dual}, and non-reciprocal acoustic devices~\cite{zhang2024nonreciprocal}. 
TISs can be realized through various approaches. One of the most classical examples is the Su–Schrieffer–Heeger model \cite{zhang2025multiple, asboth2016short}, in which topological phase transitions are driven by varying the ratio of intra- and inter-cell hopping strengths. In two-dimensional systems, TISs can be realized through valley Hall~\cite{zheng2025three, xue2024three, zhao2023topological}, quantum Hall~\cite{zhang2024observation, chen2016tunable,zhang2010topological}, and quantum spin Hall effects~\cite{chen2024elastic, hu2022simultaneous,zhang2025defect}. These mechanisms rely on breaking lattice symmetry, breaking time-reversal symmetry, and constructing pseudo-spin degrees of freedom, respectively. More recently, higher-order topological PnCs have also attracted significant attention because they can support both TISs and TCSs~\cite{ni2019observation, wang2026global, wang2022topological}.  Conventional $d$-dimensional topological insulators obey the bulk–boundary correspondence, supporting gapless $(d-1)$-dimensional boundary states. In contrast, $n$th-order topological insulators in $d$ dimensions can host $(d-n)$-dimensional states. In particular, second-order topological states in two-dimensional systems manifest as zero-dimensional corner states, offering opportunities for strong wave localization and confinement. In such topological materials, the existence of a gap associated with a TIS is generally a prerequisite for the emergence of TCSs. Zhang \textit{et al.} \cite{ZHANG2024110143} realized a valley-Hall higher-order topological phase in an elastic PnC plate by rotating scatterers to break inversion symmetry. Man \textit{et al.} \cite{man2025acoustic} designed a Lieb-latticed PnC that supports topologically protected flat bands, edge states, and higher-order corner states, and experimentally demonstrated their coexistence and localization properties. Zheng \textit{et al.} \cite{zheng2026higher} demonstrated the coexistence of higher-order TCSs and helical edge states within the same bulk BG in an acoustic spin-Chern insulator. Beyond two-dimensional systems, higher-order topological semimetals and hinge states have been demonstrated in three-dimensional acoustic crystals \cite{wei2021higher, wei20213d}. In addition, non-Hermitian acoustic systems have exhibited hybrid-order skin effects with both edge and corner localization \cite{wu2025hybrid}, as well as acoustic bound pairs induced by nonreciprocal interactions \cite{pu2026acoustic}. Despite these advances, the inverse design of PnCs with customized topological states remains a challenging task. Conventional design approaches typically rely on iterative trial-and-error procedures and extensive numerical simulations. This process usually involves repeatedly calculating numerous band structures and evaluating their topological properties to identify candidate structures. Furthermore, the verification of TISs and TCSs typically requires additional supercell calculations and eigenmode analyses, making the inverse process extremely time-consuming, particularly for complex topological states.

The rapid development of deep learning has provided a promising approach to address this challenge. By learning the complex nonlinear relationships between structural configurations and physical properties, neural networks can be employed to efficiently establish a mapping between them. Numerous studies have successfully applied deep learning techniques to inversely design PnCs with desired wave characteristics, including dispersion curves \cite{zhang2026inverse, wang2023embedding}, BG properties \cite{zhang2026accelerated, WAN2025110335, qi2025reverse}, and transmission responses \cite{zhang2026deep, huang2024inverse}. For studies involving topological acoustics, Li \textit{et al.} \cite{li2024machine} achieved acoustic edge states by designing the structural parameters using artificial neural networks, and subsequently realized two acoustic splitters with different functionalities. Chen \textit{et al.} \cite{CHEN2025110489} proposed a physics-informed neural network framework for the inverse design of low-frequency and broadband phononic metamaterials, enabling the realization of TISs for wave transport. Fan \textit{et al.} \cite{FAN2025103934} designed multiband higher-order elastic topological insulators with BGs using a deep learning framework, achieving TCSs at multiple frequencies. In addition, topology optimization based on metaheuristic algorithms, such as genetic and particle swarm algorithms, have been widely employed for the inverse design of topological PnCs with desired BGs for the generation of topological states \cite{xu2025customizing, WANG2024129571, qin2023inverse}. However, most existing studies focus on the prediction and design of BG frequencies and related topological properties, while the inverse design of TISs and TCSs remains largely unexplored. Although existing approaches can generate PnCs with nontrivial bulk BGs, they cannot directly ensure the emergence of desired TISs or TCSs at prescribed frequencies. Additional numerical simulations are often needed to identify these states and determine their frequencies. Moreover, the formation of higher-order TCSs depends on interface and boundary conditions, further increasing the complexity and computational cost of conventional design methods. Therefore, directly targeting TISs and TCSs at prescribed frequencies can reduce repeated numerical simulations and structural screening, thereby improving inverse design efficiency.

To achieve this goal, an appropriate inverse design framework is required to address the non-unique mapping between target responses and unit cell geometries. Furthermore, the inverse design of these topological states must account for hierarchical physical requirements. In particular, the design of TCSs is considerably more challenging than that of conventional topological properties, since the emergence of TCSs requires not only BGs of bulk states~\cite{xiong2025experimental} but also TIS gaps. To this end, this paper proposes a Latent-Consistency Tandem Network for the inverse design of two-dimensional topological phononic crystals. Distinct from traditional parametric design methods, this framework adopts PnC unit cells with free-form scatterers, significantly increasing the design freedom. Unlike conventional tandem networks, which rely primarily on target response matching, the proposed model incorporates latent-space consistency and structural reconstruction constraints. These constraints promote consistency between the generated latent representations and their corresponding geometric structures. This helps reduce the risk of generating physically implausible designs and improves the reliability of the inverse design process. As a result, the proposed method can not only inversely design PnCs with specified BG characteristics but also directly achieve TISs and TCSs at target frequencies. The remainder of this paper is organized as follows. The PnC models, the construction of datasets, and the proposed deep learning frameworks are discussed in Section \ref{section2}. Section \ref{section3} presents the results and discussions, including the dimensionality reduction of unit cell images, the prediction and inverse design of topological PnCs for three main objectives, as well as the numerical and experimental validation of inversely designed PnC structures. Finally, the conclusions are summarized in Section \ref{section4}.

\section{Model and methods} \label{section2}

\subsection{Model and dataset}

\begin{figure*}[htbp]
	\centering
	\includegraphics[scale=1]{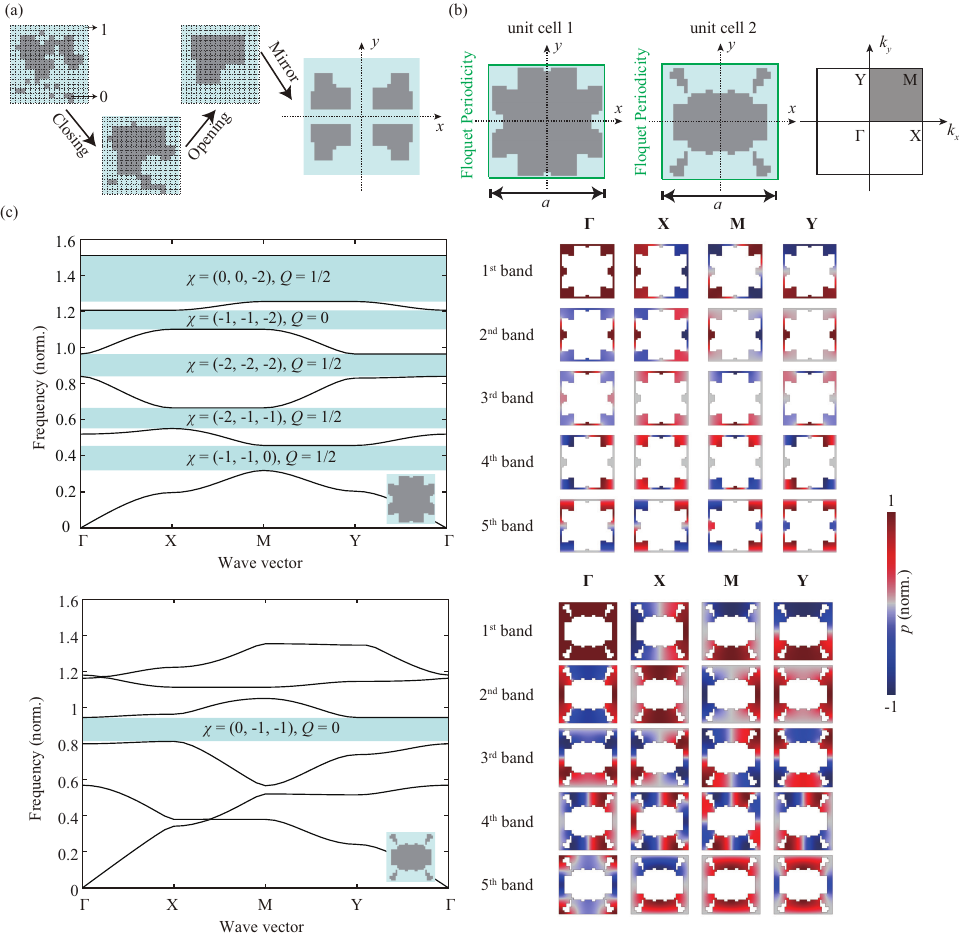}
	\caption{\label{Fig.Gener_Sample_BG}(a) Schematic illustration of the construction process for the $C_{2\textrm{v}}$-symmetry PnC unit cell, where a randomly initialized $16 \times 16$ pixel matrix is processed by morphological closing and opening operations and then expanded through mirror symmetry. The blue and gray colors represent the air background and solid scatterers, respectively.
    (b) Schematic illustrations of the two PnC unit cells (left and middle panel) and the first irreducible Brillouin zone of the square lattice (right panel).
    (c) Band structures (left panel) and eigenmodes at the high-symmetry points of the two-dimensional PnC unit cells (right panel), where the blue shaded regions denote the bulk BGs.}
\end{figure*}

In this study, a pixelated distribution of solid material, represented as a matrix, is employed to construct two-dimensional PnC unit cells made of rigid free-form scatterers embedded in air with $C_{2\textrm{v}}$ symmetry. The construction process is illustrated in Fig. \ref{Fig.Gener_Sample_BG}(a). First, a random $16 \times 16$ binary matrix is generated, where 1 and 0 represent the air (in blue) and rigid (in gray) materials, respectively.
Subsequently, morphological closing and opening operations are applied to remove isolated regions and narrow structures. Finally, mirror symmetries along the $x$- and $y$-directions are performed to obtain the final unit cell images using $32 \times 32$ pixels. In addition, to ensure the physical feasibility and periodic continuity of a generated PnC structure, the unit cell is required to satisfy the following constraints: the air region must remain simply connected and also connected to neighboring unit cells along both the $x$- and $y$-axes. Meanwhile, the number of connected scatterer regions is constrained to range from 1 to 6. 

In this paper, all numerical simulations are conducted using the finite element software COMSOL Multiphysics. Taking two PnC unit cells as examples, the first six dispersion curves are calculated by sweeping the Bloch wave vector $\mathbf{k}$ along the high-symmetry path $\Gamma$-X-M-Y-$\Gamma$ within the first irreducible Brillouin zone, as shown in Fig.~\ref{Fig.Gener_Sample_BG}(b). Bloch periodic boundary conditions are applied along both the $x$- and $y$-directions, while sound hard boundary conditions are imposed at the interfaces between the rigid scatterers and air. In the numerical simulations, the mass density and sound velocity of air are set to $\rho = 1.21$ kg/m$^{3}$ and $c = 343$ m/s, respectively. The band structures of the two shown PnC unit cells are presented in Fig. \ref{Fig.Gener_Sample_BG}(c), where the frequency is normalized by $c/a$, where $a$ is the lattice constant. Unit cell 1 exhibits the first five BGs of bulk states (namely bulk BGs), whereas unit cell 2 exhibits only the $3^{\mathrm{rd}}$ bulk BG. Importantly, the third bulk BGs of the two unit cells overlap, thereby forming a common BG.

For a two-dimensional PnC with $C_{2\textrm{v}}$ symmetry, i.e. with unit cells that are invariant under a twofold (180$^{\circ}$) rotation about the out-of-plane axis ($C_2:(x,y,z)\rightarrow(-x,-y,z)$) and under reflections with respect to two orthogonal in-plane symmetry axes ($\sigma_v:(x,y,z)\rightarrow(x,-y,z)$, $\sigma'_v:(x,y,z)\rightarrow(-x,y,z))$, the topological properties can be characterized by the following integer topological invariants: \cite{benalcazar2019quantization, van2018higher, lin2022dirac}
\begin{equation}
	[\Pi_{p}^{(2)}]=\#\Pi_{p}^{(2)}-\#\Gamma_{p}^{(2)}, \qquad \Pi=\mathrm{X},\mathrm{Y},\mathrm{M},
\end{equation}
where $\#\Pi_{p}^{(2)}$ and $\#\Gamma_{p}^{(2)}$ refers to the number of eigenmodes below the target bulk BG that exhibit the $C_2$ rotational eigenvalue $e^{2\pi i (p-1)/2} (p=1,2)$ at the high-symmetry point $\Pi$ and $\Gamma$, respectively. For the $C_{2\textrm{v}}$ symmetry considered in this work, $p=1$ and $p=2$ correspond to even- and odd-parity states, respectively. Based on this, the bulk topological index $\chi$ is defined as
\begin{equation}
	\chi^{(2)}=\left([\mathrm{X_{1}^{(2)}}],\mathrm{[Y_{1}^{(2)}]},\mathrm{[M_{1}^{(2)}]}\right),
\end{equation}
where $\mathrm{X}$, $\mathrm{Y}$, and $\mathrm{M}$ are the high-symmetry points in the first Brillouin zone. The corner topological index can be obtained from $\chi^{(2)}$ as
\begin{equation}
	Q=\frac{1}{4}\left(-\mathrm{[X_1]}-\mathrm{[Y_1]}+\mathrm{[M_1]}\right)\ \mathrm{mod}\ 1,
\end{equation}
where $Q = 1/2$ and $Q = 0$ correspond to topologically nontrivial and trivial phases, respectively. A corner topological index difference of $\Delta Q = 1/2$ indicates that TCSs can emerge at the corners of a finite structure \cite{benalcazar2019quantization}. The existence of these corner states is accompanied by TISs within the corresponding bulk BG \cite{schindler2018higher, benalcazar2017quantized}. 

\begin{figure*}[htbp]
	\centering
	\includegraphics[scale=1]{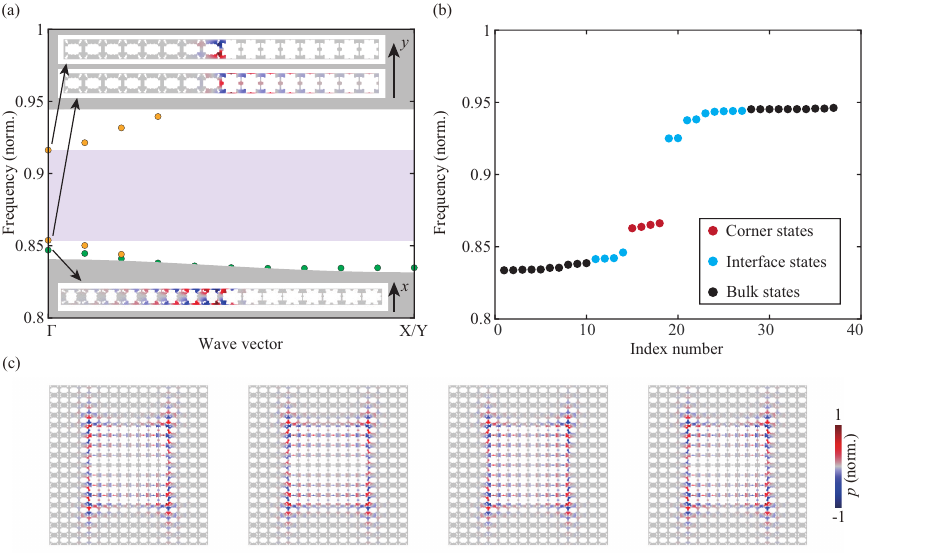}
	\caption{\label{Fig.Sample_Inter_Corner}(a) Band structures of the two PnC supercells supporting TISs propagating along the $x$- and $y$-directions, respectively. The gray shaded regions represent the bulk states, while the green and yellow dots denote the TISs propagating along the $x$- and $y$-directions, respectively. The purple shaded regions indicate the TIS gaps. The insets show the TIS eigenmodes. (b) Frequency spectrum of the PnC supercell composed of unit cells 1 and 2 supporting TCSs. (c) Acoustic pressure distributions of the TCSs indicated in (b).}
\end{figure*}

For the two selected unit cells, Fig. \ref{Fig.Gener_Sample_BG}(c) also presents the eigenmodes of the first five bands at the high-symmetry points. For each bulk BG, the bulk topological index $\chi$ and the corner topological index $Q$ are calculated from the number of the even-parity eigenmodes below the corresponding BG at different high-symmetry points. For unit cell 1, the $1^{\mathrm{st}}$, $2^{\mathrm{nd}}$, $3^{\mathrm{rd}}$, and $5^{\mathrm{th}}$ bulk BGs exhibit different values of $\chi$, but share the same nontrivial corner topological index of $Q=1/2$. Specifically, the second band exhibits even-parity at the $\Gamma$ and Y points and odd-parity at the X and M points, leading to identical changes in $\mathrm{[X_1]}$ and $\mathrm{[M_1]}$. For the third band, even-parity is observed at the $\Gamma$ and X points, whereas odd-parity occurs at the Y and M points, resulting in equal changes in $\mathrm{[Y_1]}$ and $\mathrm{[M_1]}$. These changes preserve the value of $\mathrm{[X_1]}+\mathrm{[Y_1]}-\mathrm{[M_1]}$, ensuring that the first three bulk BGs share the same nontrivial corner topological index of $Q=1/2$. This demonstrates that the corner topological index depends on the cumulative symmetry characteristics of all bands below the BG, rather than solely on the two adjacent bands.  Furthermore, according to the bulk–boundary correspondence, differences in the bulk topological indices of the two unit cells provide a theoretical basis for the emergence of TISs. For the third bulk BG, unit cells 1 and 2 exhibit $\chi=(-2,-2,-2)$ and $\chi=(0,-1,-1)$, respectively, with corresponding corner topological indices of $Q=1/2$ and $Q=0$. This topological distinction provides a basis for the emergence of TISs and higher-order TCSs when these two unit cells are joined. Two PnC supercells are constructed by combining unit cells 1 and 2, with the interfaces aligned along the $x$- and $y$-directions, respectively. The band structures corresponding to these supercells (in the form of strips) are shown in Fig. \ref{Fig.Sample_Inter_Corner}(a). The corresponding eigenmodes reveal that the acoustic pressure fields are strongly localized around the interfaces between the two kinds of PnC supercells, indicating the characteristic features of TISs. The purple shaded region denotes the common gap formed by both $x$- and $y$-directional interface states, referred to as a TIS gap. Within the frequency range of a TIS gap, the propagation of TISs is suppressed. When two interfaces along different directions are intersected to form a corner in a finite PnC structure, these overlapping TIS gaps support the emergence of zero-dimensional TCSs. Furthermore, by surrounding an $8 \times 8$ array of unit cell 1 with four layers of unit cell 2, TCSs can be realized within the PnC structure shown in Fig. \ref{Fig.Sample_Inter_Corner}(b). The corresponding eigenmodes, shown in Fig. \ref{Fig.Sample_Inter_Corner}(c), exhibit corner-localized characteristics. However, their localization is relatively weak, with part of the acoustic pressure field extending along the interfaces, because their eigenfrequencies lie near the edges of the TIS gap. Nevertheless, these eigenfrequencies remain within the TIS gap, further confirming the existence of TCSs in the proposed PnC system. Notably, by tailoring the geometric configurations (i.e., images) of PnC unit cells, a variety of acoustic properties can be realized. In this work, for the inverse design of PnCs, three objectives are primarily considered, including: (i) specific BG boundaries with target corner topological indices $Q$, (ii) designated TIS gaps, and (iii) TCSs at target frequencies. Based on the aforementioned numerical procedures, different unit cell images are randomly generated, and their corresponding band structures, topological indices, TISs, and TCSs are systematically calculated to establish the dataset for subsequent forward prediction and inverse design.

\subsection{Deep learning model}

\begin{figure*}[htbp]
	\centering
	\includegraphics[scale=1]{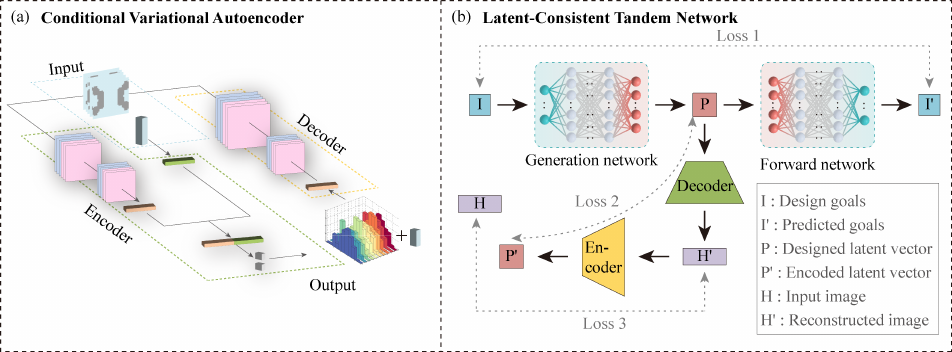}
	\caption{\label{Fig.Network}Schematic diagrams of the conditional variational autoencoder (a) and Latent-Consistent Tandem Network (b) frameworks. The Latent-Consistent Tandem Network consists of a generative network and a forward network, both of which are constructed based on the multilayer perceptrons.}
\end{figure*}

A conditional variational autoencoder is an extension of the traditional variational autoencoder \cite{kingma2013auto} that enables controllable data generation by incorporating conditional information into the generative process. In addition to the PnC unit cell images, the number of scatterers is also introduced as an input condition, allowing the scatterer number in the generated unit cells to be further regulated during the subsequent design process. Owing to the $C_{2\textrm{v}}$-symmetry, only one-quarter of the unit cell image is employed in the conditional variational autoencoder framework. Consequently, the input dimension of the network consists of a 16×16 pixel matrix together with a 1×1 conditional parameter, as illustrated in Fig.~\ref{Fig.Network}(a). The conditional parameter is the number of simply connected solid regions of scatterers in the unit cell. The loss function of the conditional variational autoencoder is composed of two parts, namely the Kullback--Leibler divergence loss and the reconstruction loss, which can be expressed as follows:
\begin{widetext}
\begin{equation}
	\begin{aligned}
		L_{\mathrm{VAE}}
		=& -\frac{1}{M}\sum_{m=1}^{M}
		\Bigg\{
		\frac{1}{2}\sum_{i=1}^{10}
		\left[
		1-(\alpha_i^{(m)})^2
		-(\sigma_i^{(m)})^2
		+2\ln\left(\sigma_i^{(m)}\right)
		\right]
		\\
		&\left.
		-\sum_{n=1}^{N}
		\left[
		x_n^{(m)}
		\ln\left(\hat{x}_n^{(m)}\right)
		+\left(1-x_n^{(m)}\right)
		\ln\left(1-\hat{x}_n^{(m)}\right)
		\right]
		\right\},
	\end{aligned}
\end{equation}
\end{widetext}
where $M$ denotes the minibatch size used during the training process. The parameters $\alpha$ and $\sigma$ correspond to the mean and standard deviations of the latent space distribution, respectively. $N$ represents the number of design variables for the original unit cell images, corresponding to a $16 \times 16$ pixel matrix. Besides, $x$ and $\widehat{x}$ denote the original and reconstructed pixel values of the unit cell, respectively. To enable stochastic sampling in the latent space, the latent vector $\mathbf{z}=[z_1,z_2,\ldots,z_I]$ is obtained through a re-parameterization that combines Gaussian noise with the learned distribution parameters:
\begin{equation}
	z_i=\alpha_i+\sigma_i\varepsilon,
	\qquad i=1,2,\ldots,I,
\end{equation}
where $\varepsilon$ is randomly sampled from a standard normal distribution. In this work, $I=10$ is chosen as a compromise between latent space compactness and reconstruction accuracy. During the training, the Adam optimization algorithm is employed with a learning rate of 0.0001 and a batch size of 512 to ensure stable convergence of the conditional variational autoencoder model. The schematic of the framework is illustrated in Fig. \ref{Fig.Network}(a).

A multilayer perceptron consists of an input layer, several hidden layers, and an output layer, and is capable of modeling complex nonlinear relationships. By constructing both generation and forward networks based on the multilayer perceptron architecture, both physical property prediction and inverse design of PnCs can be achieved.
As illustrated in Fig. \ref{Fig.Network}(b), the prediction network is employed to establish the mapping relationship between the latent vectors and various physical properties. The corresponding loss function is defined as
\begin{equation}
	\label{eq:loss_forward}
	L_{\mathrm{forward}}
	=
	\frac{1}{M}
	\sum_{i=1}^{M}
	\sum_{j=1}^{D}
	\left|
	y_{ij}-\widehat{y}_{ij}
	\right|,
\end{equation}
where $y_{ij}$ and $\widehat{y}_{ij}$ denote the real and predicted values, respectively. The dimension $D$ varies depending on the physical properties. For example, $D = 12$ for the prediction of BG boundaries, while $D = 1$ to predict $Q$.

The generation network is employed in the inverse design process to generate the latent vectors corresponding to the target physical properties. In this work, a Latent-Consistent Tandem Network is adopted, as schematically shown in Fig. \ref{Fig.Network}(b). Compared with the conventional tandem neural network, the proposed Latent-Consistent Tandem Network introduces two additional loss terms to further constrain the latent space consistency and improve the stability of the inverse design process. Accordingly, the overall loss function can be expressed as follows:
\begin{widetext}
\begin{equation}
	\label{eq:loss_inver}
	\begin{aligned}
		\textrm{Loss}
		=&
		\lambda_{1}\frac{1}{M}
		\sum_{i=1}^{M}
		\sum_{j=1}^{D}
		\left|
		y^{*}_{ij}-\widehat{y}_{ij}
		\right|
		+
		\lambda_{2}\frac{1}{M}
		\sum_{i=1}^{M}
		\sum_{j=1}^{10}
		\left|
		l_{ij}-\widehat{l}_{ij}
		\right|
		\\
		&-
		\lambda_{3}\frac{1}{M}
		\sum_{m=1}^{M}
		\sum_{n=1}^{N}
		\Big[
		x_n^{(m)}
		\ln\left(\widehat{x}_n^{(m)}\right)
		+
		\left(1-x_n^{(m)}\right)
		\ln\left(1-\widehat{x}_n^{(m)}\right)
		\Big],
	\end{aligned}
\end{equation}
\end{widetext}
where $\lambda_1$, $\lambda_2$, and $\lambda_3$ are weighting coefficients used to balance the contributions of different loss terms during network training, $y^{*}_{ij}$ denotes the target design property, while $l_{ij}$ and $\widehat{l}_{ij}$ represent the designed latent vector generated by the generation network and the reconstructed latent vector obtained from the encoder, respectively. The first term of Eq.~(\ref{eq:loss_inver}), denoted as Loss 1 in Fig. \ref{Fig.Network}(b), measures the prediction error between the target properties and the outputs of the forward network. The second term corresponds to Loss 2 and constrains the designed latent vectors to remain consistent with the latent space distribution learned by the encoder, thereby preventing the generated latent vectors from deviating excessively from the valid latent manifold. The third term, corresponding to Loss 3, is the reconstruction loss of the decoded unit cell images, which is introduced to improve the physical validity and manufacturability of the generated structures. Compared with the prediction loss, the weights of the latter two terms are relatively small, especially for the reconstruction loss. Therefore, the proposed framework does not enforce exact reconstruction of original unit cell images, but instead encourages the generated structures to remain within a physically meaningful latent space distribution. Finally, all deep learning models in this study are implemented in PyTorch \cite{paszke2019pytorch} and trained on a workstation equipped with an NVIDIA Quadro RTX 4000 (GPU) and Intel$^\circledR$ Xeon$^\circledR$ Gold 6226R (CPU). Further details of the deep learning model are available in the Appendix \ref{App2}.

\section{Results and discussions} \label{section3}

\subsection{conditional variational autoencoder results}

\begin{figure*}[htbp]
	\centering
	\includegraphics[scale=1]{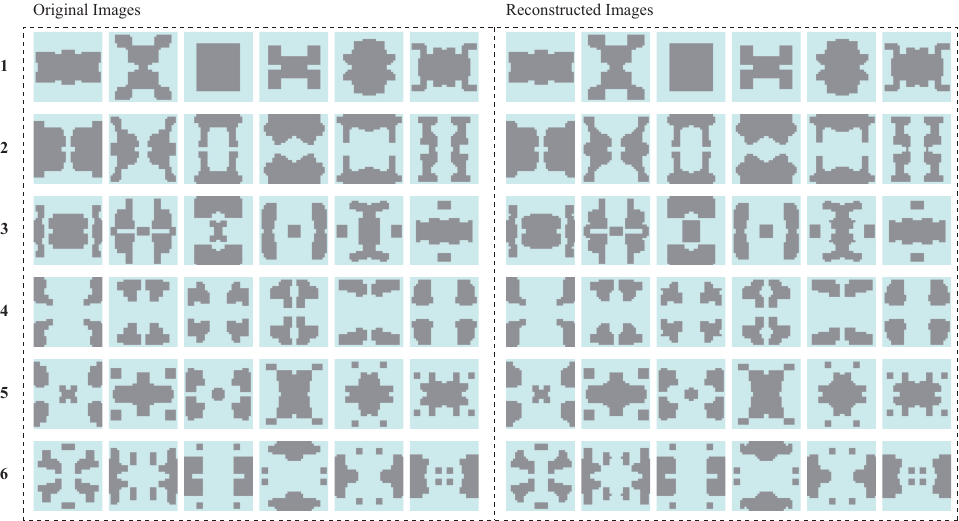}
	\caption{\label{Fig.VAE_result}Comparison between the original (left panel) and reconstructed images (right panel) of PnC unit cells from the validation set for different numbers of solid scatterers.}
\end{figure*}

During the training of the conditional variational autoencoder network, 100,000 image samples are generated for each number of solid scatterer regions ranging from 1 to 6. The entire dataset is randomly divided into a training set and a validation set with a ratio of 4:1, which are used to optimize the network parameters and evaluate the generalization ability and reconstruction performance of the trained model, respectively. To quantitatively evaluate the reconstruction capability of the network, the following formula is used to calculate the reconstruction accuracy between the original and reconstructed images:
\begin{equation}
	\label{eq:acc_VAE}
	\mathit{
		Acc_{\mathrm{VAE}}
		= \frac{1}{KN}
		\sum_{i=1}^{K}
		\sum_{j=1}^{N}
		\mathbb{I}\!\left(x_{ij} = \widehat{x}_{ij}\right)
	},
\end{equation}
where $K$ is the number of samples, and $\mathbb{I}(\cdot)$ is the indicator function, which equals 1 when $x_{ij} = \widehat{x}_{ij}$ and 0 otherwise. After the training converges, the reconstruction accuracies of the network on the training set and validation set reach 0.976 and 0.973, respectively, indicating good reconstruction capability and generalization performance of the proposed model. The pixel-level discrepancies associated with the achieved reconstruction accuracy are much smaller in scale than the acoustic wavelength considered in this study (approximately 1 m). Therefore, these minor differences are expected to have a limited influence on the overall acoustic response. To further evaluate the reconstruction quality, six samples are randomly selected from the validation set for each case with different numbers of connected solid scatterers ($1 \sim 6$), and the corresponding original and reconstructed images are shown in Fig. \ref{Fig.VAE_result}. 
We note that the reconstructed images are highly consistent with the original ones in terms of geometric features and connectivity characteristics, demonstrating that the network can effectively extract the essential features of the PnC unit cells into the latent space for dimensionality reduction while maintaining the reconstruction accuracy. It should be noted that, during the subsequent construction of the band structure dataset, the range of the latent vectors is constrained to $-5 < \mathrm{z}_{i} <5$, $i=1,2,...,I$, to avoid generating structures corresponding to excessively deviated data points in the latent space.

\subsection{Prediction results}

\begin{figure*}[htbp]
	\centering
	\includegraphics[scale=1]{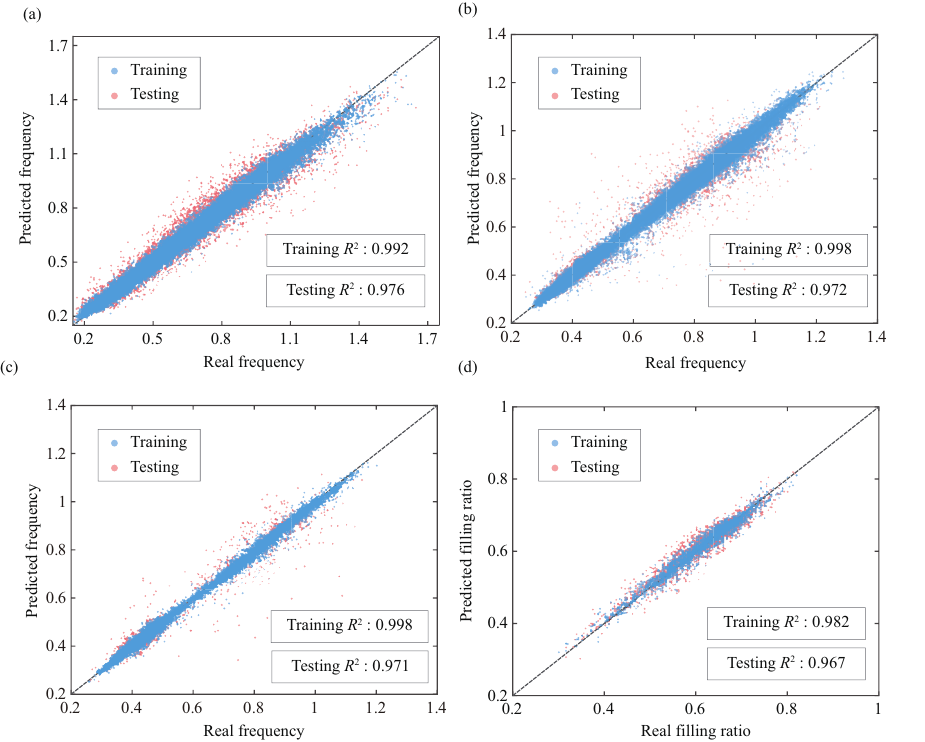}
	\caption{\label{Fig.prediction}Prediction results of the forward networks for different objectives: (a) band boundary frequencies, (b) TIS gaps, (c) TCS frequencies and (d) filling ratios of solid scatterers.}
\end{figure*}

\begin{figure*}[htbp]
	\centering
	\includegraphics[scale=0.8]{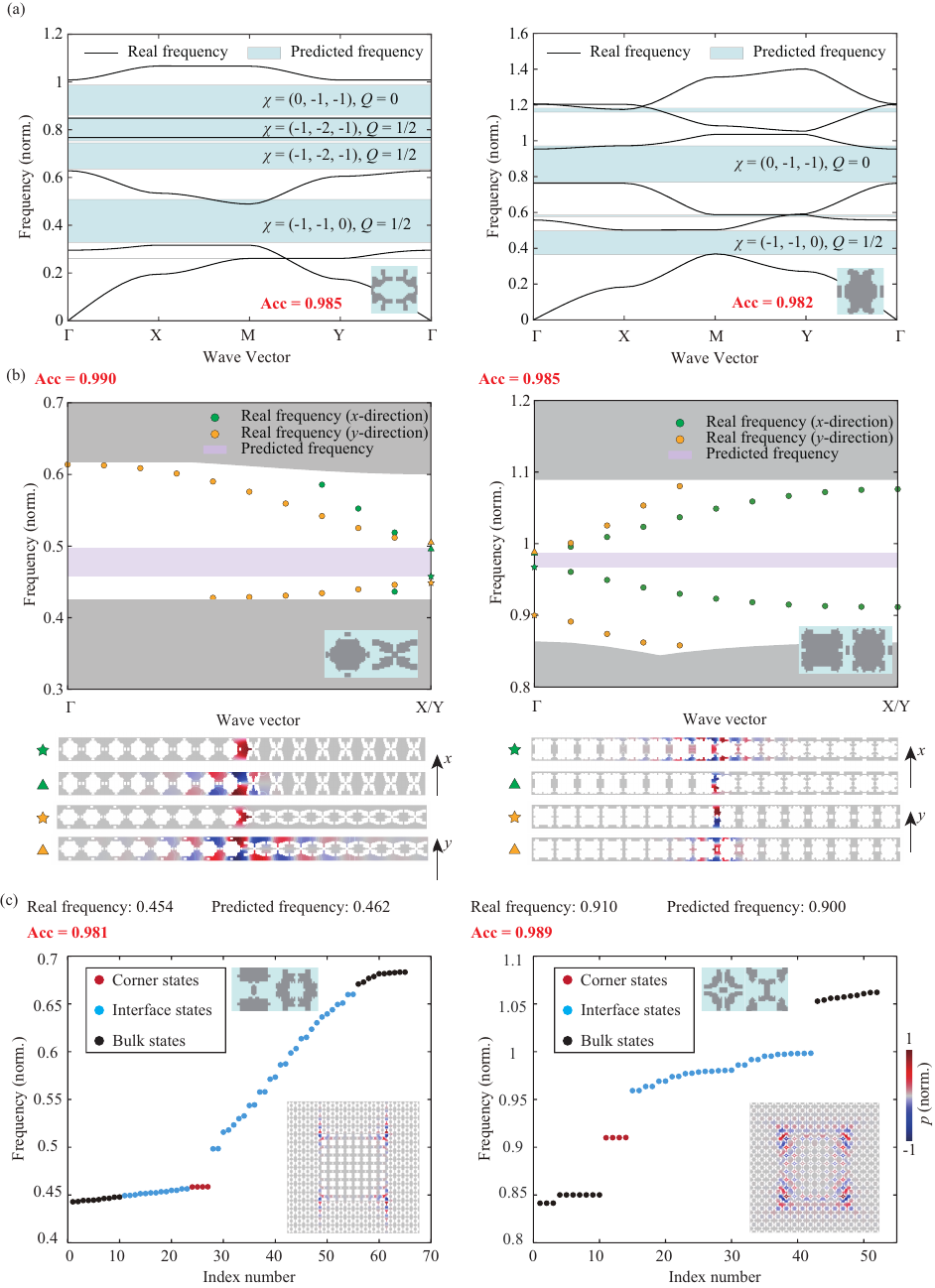}
	\caption{\label{Fig.Predicted_Sample}(a) Real band structures and predicted band boundary frequencies for two randomly selected examples from the testing set. The corresponding topological properties $\chi$ and $Q$ are also accurately predicted.
    (b) Band structures of the PnC supercells and predicted TIS gaps for two randomly selected examples. The corresponding TIS eigenmodes are also presented. The green and yellow dots denote the TISs propagating along the $x$- and $y$-directions, respectively.
    (c) Frequency spectra of the PnC supercell and predicted TCS frequencies for two randomly selected examples. The insets show the corresponding acoustic pressure distributions.}
\end{figure*}

For the first objective, the characteristic features of band structures are extracted, including the upper boundary of the first band, the upper and lower boundaries of the second to fifth bands, and the lower boundary of the sixth band. These frequencies are selected as the prediction targets because they not only determine whether the first five-order bulk BGs exist, but also define the corresponding BG frequency ranges. A total of 55,000 samples are calculated and divided into the training, validation, and testing sets with a ratio of 8:1:1. During the training process, the hyperparameters of the network are set as follows: the batch size is 128, the learning rate is 0.001, and the Adam optimizer is adopted to improve the convergence efficiency and training stability. To quantitatively evaluate the prediction performance of the network, the following correlation coefficient is employed:
\begin{equation}
	\label{eq:R2}
	\begin{split}
		R^{2} =
		\frac{
			\sum_{i=1}^{n}
			\left( y_{i}-\overline{y} \right)
			\left( \widehat{y}_{i}-\overline{\widehat{y}} \right)
		}{
			\sqrt{
				\sum_{i=1}^{n}
				\left( y_{i}-\overline{y} \right)^{2}
			}
			\sqrt{
				\sum_{i=1}^{n}
				\left( \widehat{y}_{i}-\overline{\widehat{y}} \right)^{2}
			}
		},
	\end{split}
\end{equation}
where $\overline{y}$ and $\overline{\widehat{y}}$ represent the corresponding means of the real and predicted values. The scatter distributions of the training and testing sets are presented in Fig. \ref{Fig.prediction}(a). The correlation coefficients $R^{2}$ for the training and testing sets reach 0.992 and 0.976, respectively, demonstrating that the proposed network accurately predicts the boundary frequencies of the first six bands. In addition, two examples are randomly selected from the testing set to further compare the real band structures with the predicted frequencies, as shown in Fig. \ref{Fig.Predicted_Sample}(a). The prediction accuracies for the two examples are 0.985 and 0.982, respectively. It can be observed that the proposed method accurately predicts the bulk BG boundaries. These results further verify the effectiveness and robustness of the proposed prediction framework.

From the 55,000 generated samples, the structures exhibiting the $1^{\mathrm{st}}$ to $5^{\mathrm{th}}$ BGs are further screened, and the topological property $Q$ of each BG is determined. Specifically, the data are divided into five independent datasets, where each dataset only contains structures possessing the corresponding order BG. In addition, the numbers of topologically trivial and nontrivial samples in each dataset are balanced to ensure reliable training and evaluation. Based on these datasets, separate networks are trained to predict the $Q$ values for different order bulk BGs. The Adam optimizer with a learning rate of 0.001 is adopted for all the networks. The numbers of samples in each dataset and the corresponding batch sizes are summarized in Table \ref{tab:predicted Q para}. The prediction accuracy of $Q$ is evaluated using
\begin{equation}
	\label{eq:acc_Q}
	\begin{split}
		Acc_{\mathrm{topo}}
		=
		\frac{1}{K}
		\sum_{i=1}^{K}
		\mathbb{I}
		\left(
		q_{i}=\widehat{q}_{i}
		\right),
	\end{split}
\end{equation}
where $q_{i}$ and $\widehat{q}_{i}$ denote the real and predicted topological indices, respectively, and $\mathbb{I}(\cdot)$ represents the indicator function. After the training process is completed, the classification accuracies of the training and testing sets for each order BG are presented in Table \ref{tab:predicted Q result}. Moreover, the topological properties of the BGs for the two randomly selected examples in Fig. \ref{Fig.Predicted_Sample}(a) are also accurately predicted.

\begin{table}[htbp]
	\centering
	\caption{Numbers of datasets and batch sizes for predicting the topological index \(Q\) of different order BGs.}
	\label{tab:predicted Q para}
	\begin{tabular}{c c c}
		\hline
		Order of BG & Dataset size & Batch size \\
		\hline
		\(1^{\mathrm{st}}\) BG & 2,400 & 8 \\
		\(2^{\mathrm{nd}}\) BG & 2,400 & 8 \\
		\(3^{\mathrm{rd}}\) BG & 22,910 & 512 \\
		\(4^{\mathrm{th}}\) BG & 26,000 & 512 \\
		\(5^{\mathrm{th}}\) BG & 27,791 & 256 \\
		\hline
	\end{tabular}
\end{table}

\begin{table}[htbp]
	\centering
	\caption{Prediction accuracies of the topological property $Q$ for different order BGs.}
	\label{tab:predicted Q result}
	\begin{tabular}{c c c c c c}
		\hline
		& \(1^{\mathrm{st}}\) BG & \(2^{\mathrm{nd}}\) BG & \(3^{\mathrm{rd}}\) BG & \(4^{\mathrm{th}}\) BG & \(5^{\mathrm{th}}\) BG \\
		\hline
		Training set  & 1.000 & 0.998 & 0.984 & 0.981 & 0.982 \\
		Testing set & 0.989 & 0.972 & 0.956 & 0.971 & 0.969 \\
		\hline
	\end{tabular}
\end{table}

Based on the previously established datasets, the PnC samples with identical frequency ranges but different topological properties of bulk BGs are selected to calculate the TISs along the $x$- and $y$-directions. Thus, the TIS gaps are obtained, which serve as the second objective. It should be noted that the BG order is no longer emphasized in this stage, and only the BG frequency range is considered. Finally, a total of 39,130 supercell samples are calculated for the prediction of TIS gaps. Different from the previous prediction task, the latent vectors corresponding to two unit cells are simultaneously used as the inputs of the network in this stage. The prediction results of the training and testing sets are presented in Fig. \ref{Fig.prediction}(b). The correlation coefficients $R^{2}$ for the training and testing sets reach 0.998 and 0.972, respectively, indicating excellent prediction capability of the proposed network for the TIS gaps. Furthermore, two PnC samples are randomly selected from the testing set, and the corresponding dispersion curves of TISs and the prediction results are presented in Fig. \ref{Fig.Predicted_Sample}(b). Additionally, the eigenmodes corresponding to the TIS propagating along both the $x$- and $y$-directions at the TIS gap boundaries are displayed. All of these modes exhibit pronounced TIS characteristics, with the acoustic energy highly localized at the interfaces. The numerically calculated results are consistent with the predicted TIS gaps.

Next, the TCSs of PnC structures are calculated for the third objective. According to the previously obtained TIS gaps, the approximate frequency ranges of corner states can be determined in advance, which significantly reduces the computational cost and simulation time. The TCS samples are selected only from the structures with confirmed TIS gaps. In addition, the corner topological indices of the two constituent unit cells are calculated. This further distinguishes the TCSs from trivial defect states. The corresponding acoustic pressure fields are further analyzed to confirm the spatial localization characteristics of the corner states. A total of 17,390 samples are established for the prediction task. The prediction results are presented in Fig. \ref{Fig.prediction}(c), where the $R^{2}$ of the training and testing sets reach 0.998 and 0.971, respectively. Similarly, two examples are randomly selected from the testing set, as shown in Fig. \ref{Fig.Predicted_Sample}(c). It can be observed that the proposed method accurately predicts the TCS frequencies, and the corresponding acoustic pressure fields are highly localized at the corners.

For the prediction of the three objectives described above, the inputs of networks include not only the latent vectors, but also the filling ratios and the numbers of scatterers. Since the filling ratio of a unit cell is also considered as one of the targets in the inverse design process, an additional network is separately established to predict the mapping relationship between the latent vectors and the filling ratios. A total of 100,000 samples are generated and divided into the training, validation, and testing sets for network training and evaluation, in a ratio of 8:1:1. The prediction results are presented in Fig. \ref{Fig.prediction}(d), demonstrating that the proposed network accurately captures the relationship between the latent representations and the corresponding filling ratios, thereby providing an effective constraint for the subsequent inverse design process.

To further evaluate the prediction efficiency of the proposed framework, Table \ref{tab:running time} compares the average prediction time of the proposed deep learning models with the corresponding finite element simulation time for each objective. For each objective, the prediction time is evaluated by averaging the computational time required to process 1000 samples over three independent runs. The results demonstrate that the proposed framework significantly improves computational efficiency. For all objectives, the prediction time is on the order of microseconds, whereas the corresponding finite element simulations require several minutes. In particular, for the prediction of TCS frequencies, the computational speed is more than $1.0\times10^{7}$ times faster than conventional finite element simulations. In contrast, the proposed deep learning models can rapidly predict the target properties of a given PnC structure, providing an efficient surrogate model instead of time-consuming numerical simulations.

\begin{table}[h]
	\centering
	\caption{Average running time required for deep learning predictions and corresponding finite element (FE) simulations}\label{tab:running time}
	\begin{tabular}{cccc}
		\hline
		Design target & Prediction ($\mu$s) & FE simulations (min) \\
		\hline
		BG boundaries  &  47.9   & 1.5  \\
		$Q$  &  53.9   & 1.5  \\
		TISs & 49.8  & 2.5  \\
		TCSs &  81.5  & 13.5  \\
		\hline
	\end{tabular}
\end{table}

\subsection{Inverse design}

\begin{figure*}[htbp]
	\centering
	\includegraphics[scale=1]{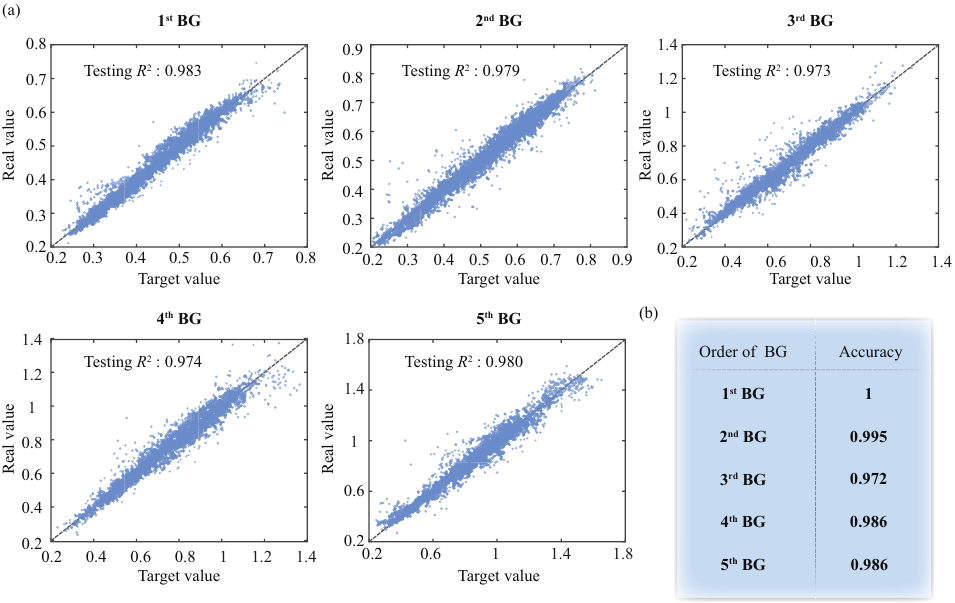}
	\caption{\label{Fig.BG_Inver}Design results of the Latent-Consistent Tandem Network for different objectives: (a) bulk BG boundary frequencies and (b) topological index $Q$.}
\end{figure*}

For the first objective, the inverse design aims to generate PnC unit cell images with target BG boundary frequencies and desired topological index $Q$. Different BG orders are designed using separate datasets, whose sample sizes are summarized in Table \ref{tab:predicted Q para}. Apart from the physical properties, the filling ratio of solid scatterers is also incorporated as a design objective to improve the manufacturability and geometric controllability of the generated PnC structures. Therefore, for this inverse design task, the dimension $D$ in the first term of Eq. (\ref{eq:loss_inver}) is set to 4, corresponding to the lower and upper boundary frequencies of the target BG, the topological index $Q$, and the filling ratio, respectively. This multi-objective formulation enables the network to simultaneously achieve accurate control of the frequency and topological characteristics of the targeted BGs, and the structural compactness of the designed unit cells. The scatter plots of target and real values, including the BG boundary frequencies and filling ratios, for different BG orders in the testing sets are shown in Fig. \ref{Fig.BG_Inver}(a), while the accuracy of topological index $Q$ is presented in Fig. \ref{Fig.BG_Inver}(b). The real values are obtained through the numerical simulations using the PnC unit cells reconstructed by decoding the designed latent vectors. The correlation coefficient for each order BG exceeds 0.97, indicating that the proposed networks can achieve accurate inverse design of the target properties. Additionally, different physical properties are specified for each order BG to investigate the variations in the generated PnC unit cells. The results for the $1^{\mathrm{st}}$ BG are shown in Fig. \ref{Fig.Inver_BG1_Samples}, while the results for the other four orders of BGs are provided in the Appendix \ref{App1}. For different BG boundaries, as well as different filling ratios, the networks can successfully generate unit cell images satisfying the target requirements. Meanwhile, it can be observed that the geometric morphologies of the generated structures vary significantly with the BG frequency range and filling ratio, demonstrating that the proposed method can effectively learn the mapping relationship between the geometric configurations of unit cells and their physical properties. For the cases with different topological indices $Q$ under the same BG range, the generated unit cells also exhibit significantly different structural characteristics. Figure \ref{Fig.Inver_BG1_Samples}(b) presents the band structure corresponding to a designed PnC unit cell. The eigenmodes of the first band at the high-symmetry points are also provided. The parities of these eigenmodes further confirms that the designed BG possesses a topological index of $Q=1/2$. Overall, it is demonstrated that the established networks can accurately achieve the inverse design of unit cells with target BG boundaries, filling ratios, and topological properties simultaneously.

\begin{figure}[htbp]
	\centering
	\includegraphics[scale=1]{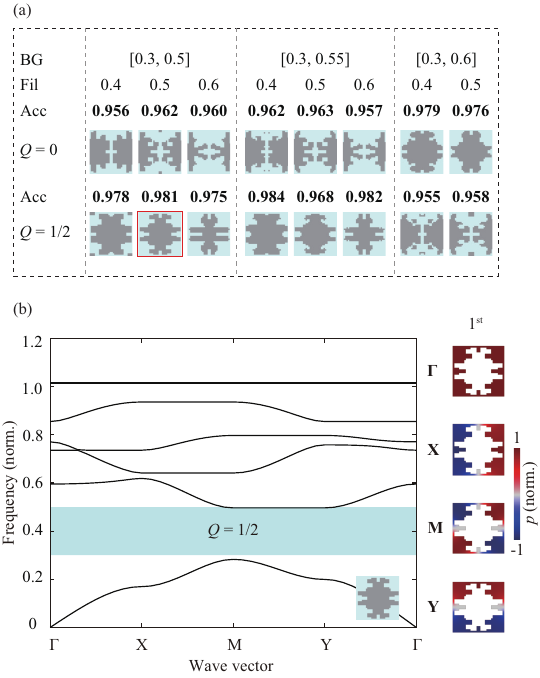}
	\caption{\label{Fig.Inver_BG1_Samples}(a) Representative generated unit cells for the inverse design of the $1^{\mathrm{st}}$ BG under different target BG boundaries, filling ratios, and topological indices, together with the corresponding design accuracies. (b) Band structure and eigenmodes at the high-symmetry points of the unit cell highlighted in (a). The blue shaded region indicates the target BG range.}
\end{figure}

\begin{figure*}[htbp]
	\centering
	\includegraphics[scale=1]{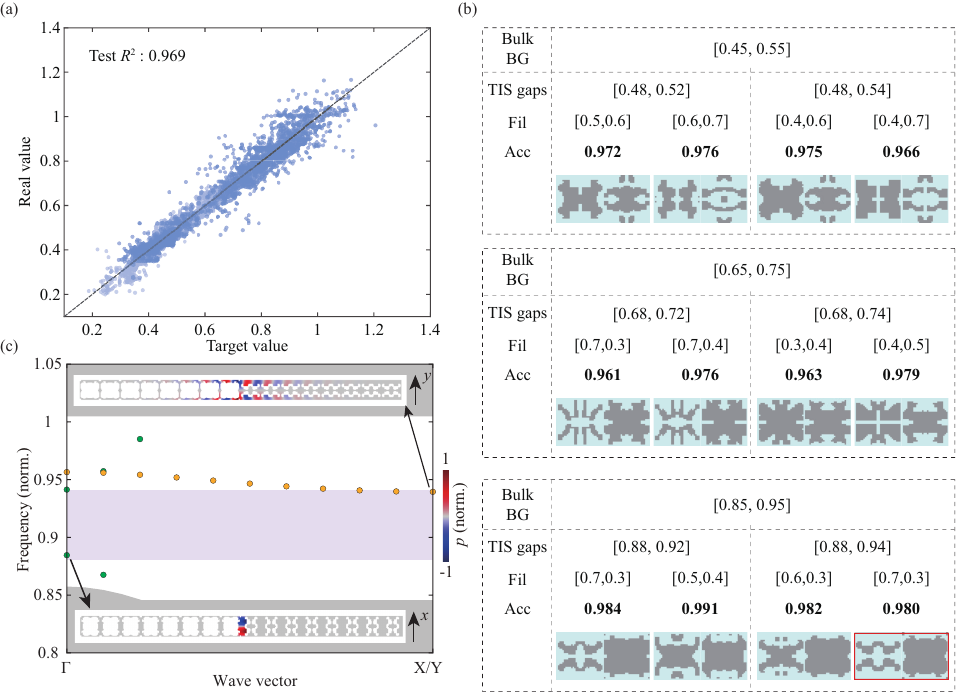}
	\caption{\label{Fig.Inver_TIS_Samples}(a) Scatter plot comparing the designed and target TIS gaps in the testing set. (b) Representative unit cell pairs generated under different bulk BG boundaries, target TIS gaps, and filling ratio constraints, together with the corresponding design accuracies. (c) The band structure and eigenmodes of the PnC supercells composed of representative unit cell pair highlighted in (b). The green and yellow dots are the TISs propagating along the $x$- and $y$-directions, respectively. The purple shaded region represents the target TIS gap range.}
\end{figure*}

The dataset used for the inverse design of TIS gaps is identical to that employed in the prediction stage. For this design objective, the filling ratio of solid scatterers in a unit cell is also included as a target constraints, thereby enabling multi-objective inverse design. Additionally, the bulk BG boundaries are incorporated as input parameters during the inverse design process. To better satisfy the physical requirements of this task, the first term in Eq. (\ref{eq:loss_inver}) is reformulated by incorporating additional constraints. Specifically, besides matching the target TIS gap and filling ratio, the generated unit cell pairs are further constrained to exhibit the desired bulk BG boundaries and topological properties. These constraints ensure that the designed unit cell pairs share the same frequency range of bulk BGs while possess distinct topological characteristics. It is worth noting that the BG order is not explicitly constrained, enabling the networks to automatically discover unit cell pairs corresponding to different BG orders that satisfy the prescribed wave characteristics. After training, the inverse design results for the testing set are shown in Fig. \ref{Fig.Inver_TIS_Samples}(a). A correlation coefficient of 0.969 is achieved, demonstrating the high accuracy of the proposed inverse design framework. Figure \ref{Fig.Inver_TIS_Samples}(b) presents several generated unit cell pairs under different design targets, together with the corresponding design accuracies. For identical bulk BG boundaries, unit cells with different target TIS gaps are successfully generated. Furthermore, the filling ratios and numbers of scatterers can be specified, indicating the diversity of design solutions. Importantly, both low-frequency and high-frequency TIS gaps can be accurately achieved under different bulk BG boundary conditions, further demonstrating the robustness and flexibility of the proposed method. To validate the designed structures, a unit cell pair is selected for analysis, where the band diagrams of the corresponding PnC supercells are shown in Fig. \ref{Fig.Inver_TIS_Samples}(c). By examining the corresponding eigenmodes in both band diagrams, the TIS gaps can be clearly identified, showing good agreement with the target design. These results further verify that the proposed inverse design framework can accurately generate the pairs of PnC unit cells with the target TIS gap characteristics while maintaining the required topological properties.

\begin{figure*}[htbp]
	\centering
	\includegraphics[scale=1]{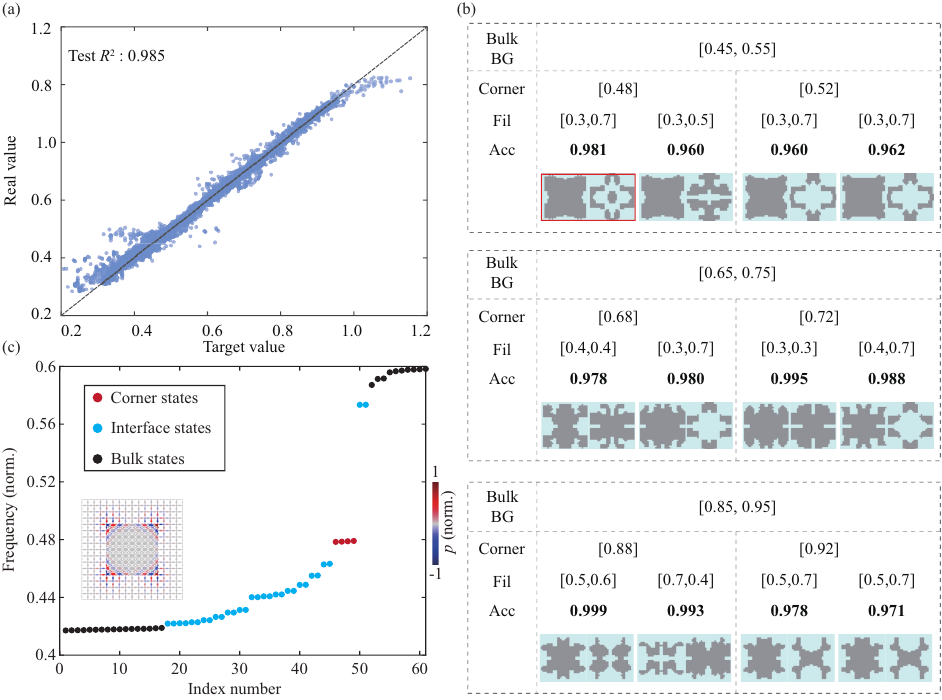}
	\caption{\label{Fig.Inver_Corner_Samples}(a) Scatter plot comparing the designed and target TCS frequencies in the testing set. (b) Representative unit cell pairs generated under different bulk BG boundaries, target TCS frequencies, and filling ratio constraints, together with the corresponding design accuracies. (c) Frequency spectrum of the PnC supercell composed of the representative unit cell pair highlighted in (b). The insets show the acoustic pressure distribution of the TCS.}
\end{figure*}

For the inverse design of TCSs, the same dataset as that used in the corresponding prediction stage is employed. To enable multi-objective inverse design, the filling ratio is incorporated as an additional design constraint together with the target TCS frequency. The loss function is constructed in the same manner as that used for the TIS design task, where the target frequency, filling ratio, and topological properties are jointly considered to guide the generation of corresponding PnC structures. The inverse design results for the testing set are presented in Fig. \ref{Fig.Inver_Corner_Samples}(a), where a correlation coefficient of 0.985 is achieved, demonstrating the high accuracy of the proposed inverse design framework. Figure \ref{Fig.Inver_Corner_Samples}(b) shows several representative design examples together with their corresponding accuracies. For these designed supercells, the bulk BGs and corner topological indices $Q$ of the two constituent unit cells are further verified. Moreover, the corresponding TCS frequencies are confirmed to lie within the TIS gaps. It can be observed that the trained network is capable of generating the unit cell pairs that support a wide range of target TCS frequencies. This implies that the network successfully learns the complex nonlinear relationship between the unit cell images and TCS frequencies. To further verify the physical correctness of the designed PnC structures, the eigenfrequency spectrum of one selected design is shown in Fig. \ref{Fig.Inver_Corner_Samples}(c). We note that the designed corner state lies within the TIS gap and is isolated from the bulk and interface states. The corresponding acoustic pressure distribution reveals strong energy localization at the corners, confirming the existence of a TCS. 

\begin{figure*}[htbp]
	\centering
	\includegraphics[scale=1]{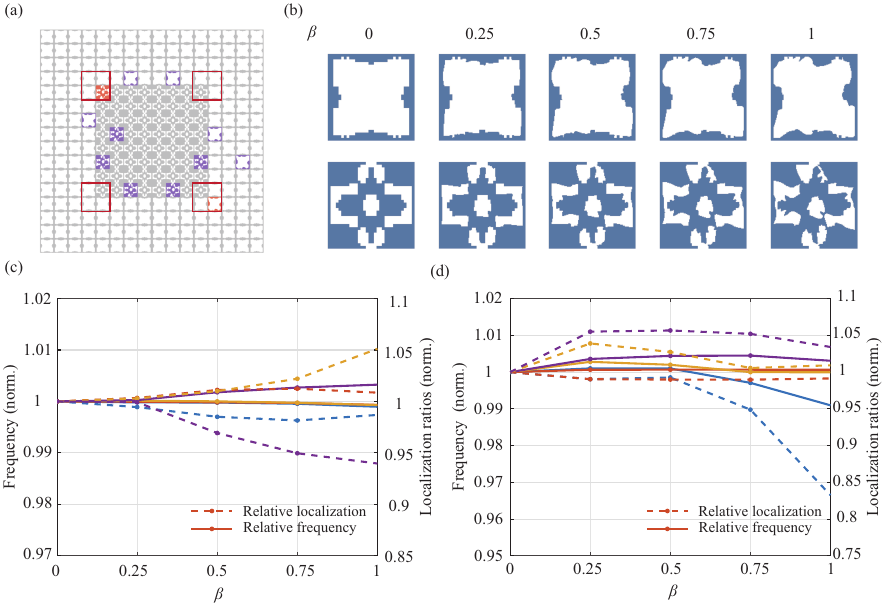}
	\caption{\label{Fig.robustness}(a) PnC supercell structure with perturbed unit cells. (b) Perturbed air domains under different perturbation strengths $\beta$. (c,d) Normalized TCS frequencies and normalized localization ratios for perturbations applied away from and near the corner regions, respectively.}
\end{figure*}

To evaluate the robustness of the designed corner states against local structural imperfections, geometric perturbations are introduced into different unit cells of the PnC supercell. Specifically, unit cells near and far from the corner region are selected [highlighted by orange and purple colors in Fig. \ref{Fig.robustness}(a), respectively], and identical continuous geometric perturbations are applied to these cells. The perturbed air domain is defined as
\begin{equation}
	\Omega_{\mathrm{air}}(\beta)=\left\{\mathbf{x}: \phi_0(\mathbf{x})+\beta d_{\max}R_c(\mathbf{x})>0\right\},
\end{equation}
where $\phi_0(\mathbf{x})=d_{\mathrm{air}}(\mathbf{x})-d_{\mathrm{solid}}(\mathbf{x})$ is the signed distance field, and $d_{\mathrm{air}}(\mathbf{x})$ and $d_{\mathrm{solid}}(\mathbf{x})$ represent the distances from a point to the air and solid regions, respectively. The function $R_c(\mathbf{x})$ is a normalized smooth random field that controls the local direction and magnitude of the boundary deformation. The dimensionless parameter $\beta$, which controls the perturbation amplitude, is set to 0, 0.25, 0.5, 0.75, and 1. Throughout this process, $R_c(\mathbf{x})$ remains unchanged across different values of $\beta$, allowing the perturbation strength to be gradually increased under a constant deformation pattern. Additionally, the outer interface regions connecting adjacent cells remain fixed to preserve periodic connectivity, while the internal boundaries are allowed to deform without enforcing geometric symmetry. Figure \ref{Fig.robustness}(b) shows the evolution of the air domains in the two unit cells under different perturbation strengths. Although the air domains are continuously deformed, their overall geometric features and connections with adjacent cells are preserved.

The changes in the TCS frequencies and spatial localization characteristics of the supercell structures are then evaluated for different perturbation strengths and perturbation locations. The spatial localization ratio is quantified by calculating the ratio between the acoustic energy within the 16 unit cells enclosed by the red boxes in Fig. \ref{Fig.robustness}(a) and the total acoustic energy of the supercell. Figure \ref{Fig.robustness}(c) presents the results for the perturbations introduced in the purple unit cells away from the corner regions. The solid and dashed lines represent the normalized TCS frequencies and normalized localization ratios, respectively. Here, the normalized frequency and localization ratio are defined as the ratio of their corresponding values in the perturbed supercell to those in the unperturbed supercell, respectively. The results show that the frequencies and spatial localizations of the TCSs are well maintained under these perturbations, with only minor variations. Figure \ref{Fig.robustness}(d) presents the results for the perturbations introduced in the orange unit cells near the corner regions. Since these perturbations are introduced near the corner regions, relatively larger variations in the spatial localization are observed. However, the TCS frequencies exhibit only minor fluctuations, and the acoustic fields remain confined within the designed corner regions without notable delocalization or loss. These results demonstrate that the TCSs designed using the proposed approach possess high robustness against local geometric imperfections.

At last, we consider the inverse design efficiency of the deep learning frameworks. The average running time of the inverse design for each objective is summarized in Table \ref{tab:running time inver}. The time is obtained from their independent tests, each involving the generation of 1,000 PnCs. Traditional design approaches generally require extensive numerical simulations and iterative screening of candidate structures, whereas the proposed framework completes each inverse design within only a few tens of microseconds. This further demonstrates the design efficiency of the proposed framework.

\begin{table}[h]
	\centering
	\caption{Average running time required for the inverse design}\label{tab:running time inver}
	\begin{tabular}{cc}
		\hline
		Design target & Latent-Consistent Tandem Network (ms)\\
		\hline
		BG properties  &  1.3   \\
		TISs & 1.4   \\
		TCSs &  6.1  \\
		\hline
	\end{tabular}
\end{table}

\subsection{Experimental validation}

\begin{figure*}[htbp]
	\centering
	\includegraphics[scale=1]{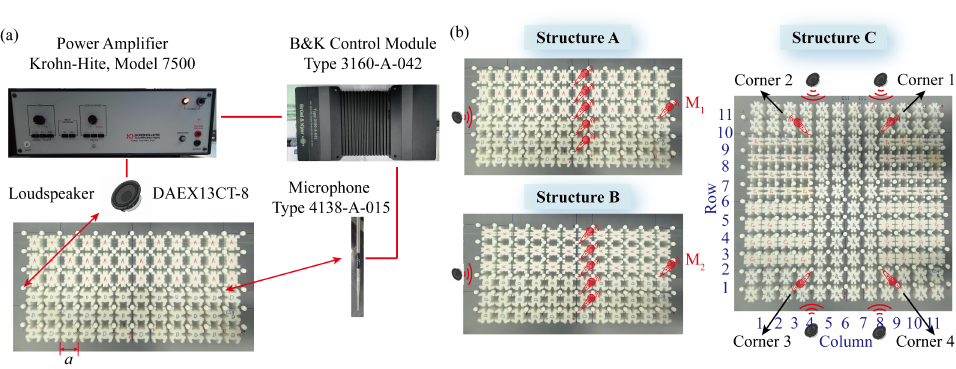}
	\caption{\label{Fig.experiment_setup}(a) Experimental setup for the measurement of acoustic waves. (b) Fabricated PnC structures used for experimental validation, where the locations of the speaker and microphone are indicated by the corresponding symbols.}
\end{figure*}

To experimentally validate the target TISs and TCSs, the PnC structures are fabricated by 3D printing according to the unit cells generated through the inverse design process. The lattice constant and height of the fabricated PnCs are 83.3 mm and 30 mm, respectively. Two PMMA plates are placed on top and bottom of the PnC samples to satisfy the two-dimensional assumption. Since the sample height is $h=30$ mm, the two-dimensional approximation remains valid for frequencies below $f=c/2h\simeq 5700$ Hz, ensuring that no higher-order modes are excited along the thickness direction. The experimental setup is shown in Fig. \ref{Fig.experiment_setup}. A loudspeaker DAEX13CT-8 is used as the acoustic source, and a microphone Type 4138-A-015 from B\&K is employed to measure the acoustic pressure.  To enable acoustic signal acquisition, periodic holes are drilled in the top PMMA plate at the positions corresponding to the gaps between adjacent PnC unit cells, into which the microphone could be inserted. During each measurement, only the acoustic signal in the hole at the target location is acquired, while all the other holes are sealed with matching plugs. Sinusoidal excitation signals are generated and post-processed using the control module (B$\&$K 3160-A-042). Note that different target functionalities are verified using the PnC structures assembled from the designed unit cells with different arrangements. As shown in Fig. \ref{Fig.experiment_setup}(b), Structures A and B are employed to verify the interface states, while Structure C is used to validate the corner states. Structures A and B both comprise $12 \times 6$ unit cells. However, their arrangement directions are different, enabling the verification of TIS propagation along different directions (corresponding to the $x$- or $y$-directional interface). Structure C possesses four regions of corner states. Each region is formed by a $4 \times 4$ of nontrivial unit cells surrounded by two layers of trivial unit cells.

\begin{figure*}[htbp]
	\centering
	\includegraphics[scale=1]{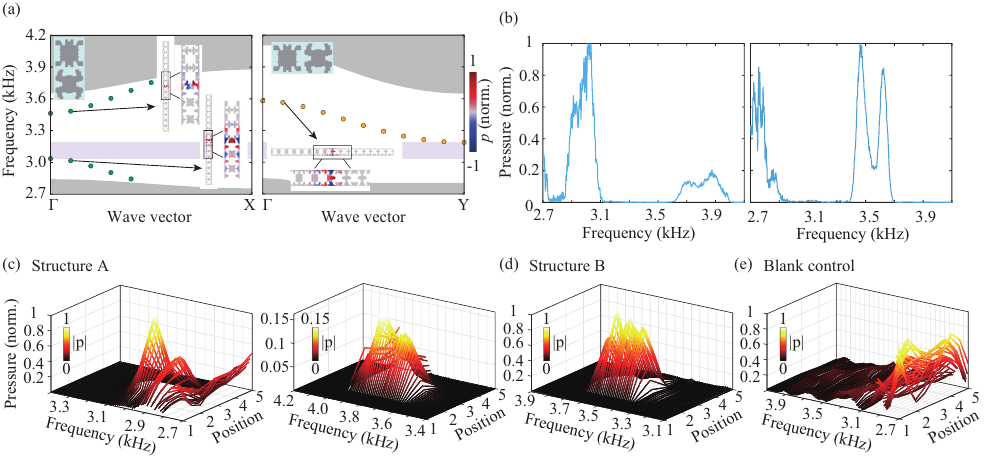}
	\caption{\label{Fig.experiment_num_TIS}(a) Numerically calculated TIS spectra along the $x$- (left panel) and $y$-directions (right panel). The gray shaded regions denote the bulk states, while the green and yellow dots represent the TISs propagating along the $x$- and $y$-directions, respectively. The insets show the eigenmodes of TISs. (b) Measured acoustic pressure spectra for Structures A (left panel) and B (right panel) at frequencies of the interface modes at the locations shown in Fig.~\ref{Fig.experiment_setup}(b) for each structure. (c), (d) and (e) present the measured acoustic pressure distributions along the interfaces for Structure A, Structure B and the blank control, respectively as shown in Fig.~\ref{Fig.experiment_setup}(b).}
\end{figure*}

A pair of PnC unit cells is designed to achieve a target TIS gap from 3.09 to 3.17 kHz. The numerically calculated band structures are plotted in the Fig. \ref{Fig.experiment_num_TIS}(a). The dispersion curves show that the TISs emerge in the frequency ranges of $2.84 \sim 3.04$ kHz and $3.45 \sim 3.75$ kHz along the $x$-direction, and $3.19 \sim 3.58$ kHz along the $y$-direction. Consequently, a common TIS gap of $3.04 \sim 3.19$ kHz is formed, corresponding to a design accuracy of 0.989. In the experiments, the unit cells are arranged along different directions to construct Structures A and B, as shown in Fig. \ref{Fig.experiment_setup}(b). The acoustic pressure is measured at positions $\mathrm{M}_{1}$ and $\mathrm{M}_{2}$, and the corresponding transmission spectra are presented in Fig. \ref{Fig.experiment_num_TIS}(b). For Structure A, two distinct transmission bands are identified, whereas Structure B exhibits a single transmission band. The high-frequency transmission band of Structure A exhibits a lower peak amplitude, possibly owing to stronger dissipation and scattering at higher frequencies. To further verify the localization characteristics of the TISs, the acoustic pressure distributions in the central regions of the two PnC structures are measured at the positions indicated in Fig. \ref{Fig.experiment_setup}(b). The normalized acoustic pressure distributions are shown in Figs. \ref{Fig.experiment_num_TIS}(c) and \ref{Fig.experiment_num_TIS}(d). In all transmission bands, significant acoustic pressure enhancement occurs near the interfaces. This behavior agrees well with the finite element simulations and further confirms the existence of the TISs. Notably, a slight frequency shift is observed between the experimental and numerical results, with the experimentally measured TIS frequencies being slightly higher than the numerical values. These discrepancies are mainly attributed to differences between the material properties assumed in the simulations and those of the actual air, together with unavoidable deviations between the idealized acoustic boundary conditions and the real experimental conditions. For comparison, the acoustic pressure is measured at the same locations within a blank waveguide without any PnCs as in the blank control experiment. The measured acoustic pressure distribution is plotted in Fig. \ref{Fig.experiment_num_TIS}(e). In contrast to the TISs, the acoustic pressure fields of the blank structure exhibits no evident localization. This comparison indicates that the observed wave transmission along the interfaces originates from the TISs rather than the excitation source.

\begin{figure*}[htbp]
	\centering
	\includegraphics[scale=1]{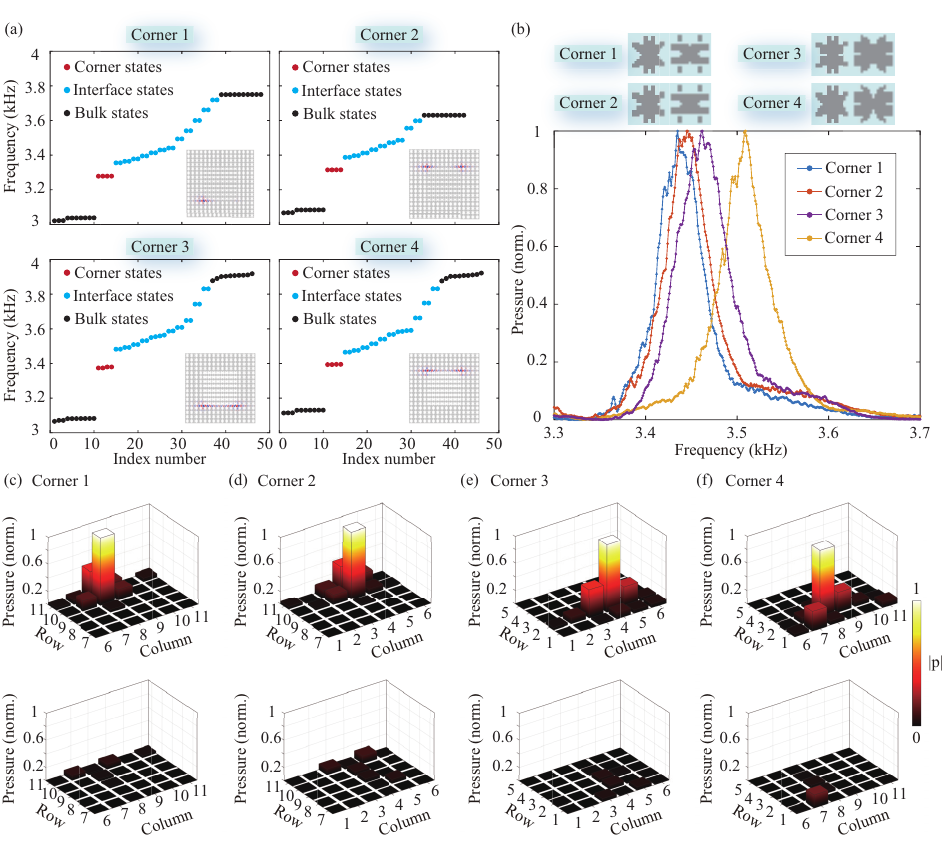}
	\caption{\label{Fig.experiment_num_Corner}(a) Frequency spectra of the four PnC supercells. (b) The four unit cell pairs used to construct the corresponding supercells and the measured acoustic pressure transmission spectra at the four corner positions. (c)–(f) Measured acoustic pressure distributions for Corner 1-4. The upper row corresponds to the TCS frequencies, and the lower row corresponds to 3.60 kHz (within the TIS gap).}
\end{figure*}

To validate the design ability of TCSs, four pairs of PnC unit cells are designed with target TCS frequencies of 3.25, 3.33, 3.42, and 3.50 kHz, respectively. The eigenfrequency spectra of the corresponding finite PnC structures are calculated and shown in Fig. \ref{Fig.experiment_num_Corner}(a). The numerically calculated TCS frequencies are 3.28, 3.32, 3.37 and 3.39 kHz, which correspond to the design accuracies of 0.991, 0.997, 0.985, and 0.969, respectively. Figure \ref{Fig.experiment_num_Corner}(b) shows the measured transmission spectra of acoustic pressure with the microphone positioned at the corner locations indicated in Fig. \ref{Fig.experiment_setup}(b). Four pronounced transmission peaks emerge in the spectra, confirming the existence of the corner states. The experimentally measured TCS frequencies show good agreement with the numerical results. To further verify the spatial localization characteristics of the corner states, the acoustic pressure fields at the four peak frequencies are measured, as displayed in Fig. \ref{Fig.experiment_num_Corner}(c). It is seen that the acoustic pressure is strongly localized at the corresponding corner positions, which coincide with the microphone locations marked in Fig. \ref{Fig.experiment_setup}(b). These locations exhibit the highest acoustic pressure amplitudes within the entire PnC structure, indicating that the acoustic energy is highly concentrated near the corners. This observation is in good agreement with the numerically results of the corner states. In addition, the pressure field distributions of all PnC structures at 3.60 kHz (in the BGs) are also presented. Compared with the pressure fields at the TCS frequencies, the acoustic pressure at this frequency is significantly weaker, and no obvious wave localization can be observed. Similar to the TISs, the experimentally measured TCS frequencies are slightly higher than the numerical predictions. In summary, the experimental results verify that the inversely designed PnCs possess the desired TIS and TCS properties.

\section{Conclusion} \label{section4}

In this study, a deep learning framework is developed for the inverse design of two-dimensional topological PnCs. The proposed framework directly customizes desired topological states of acoustic waves through free-form unit cell images with a high degree of design freedom. The conditional variational autoencoder is first employed to obtain low-dimensional representations of unit cell images, significantly reducing the complexity of inverse design in a high-dimensional design space. Based on the extracted latent vectors, the forward and generation networks based on multilayer perceptrons are developed for the accurate prediction and inverse design of BG characteristics, together with TIS and TCS frequencies. The proposed forward prediction networks significantly reduce the computational cost, achieving speedups of up to approximately $10^{7}$ times compared with conventional finite element simulations. 

Compared with conventional BG design, the inverse design of TISs and TCSs involves greater nonlinearity and complexity. Therefore, an improved Latent-Consistent Tandem Network is proposed by incorporating additional constraints into the loss function. Moreover, the framework supports multi-objective inverse design by incorporating the filling ratio as an additional design objective, enabling the simultaneous control of topological characteristics and geometric features. As a result, correlation coefficients above 0.96 are achieved for all design objective. In addition, the inverse design process can be completed on a microsecond time scale, highlighting the high computational efficiency of the proposed framework. Furthermore, customized inverse designs are successfully achieved for prescribed TIS gap ranges and target TCS frequencies. The designed PnC structures are further fabricated using 3D printing and experimentally validated. The measured transmission spectra and acoustic pressure distributions are in excellent agreement with the numerical predictions, demonstrating the successful realization of highly localized TISs and TCSs. It is worth emphasizing that this work goes beyond the conventional focus on BG characteristics and elevates the design target from the BG itself to TISs and TCSs. 

The proposed framework eliminates the need to first design BGs and then indirectly construct the corresponding topological states, thereby significantly improving the efficiency and flexibility of topological PnC design. Owing to the excellent localization and robustness of topological states, the proposed method shows great potential for applications in acoustic energy harvesting, wave filtering, and acoustic sensing. In addition, by adjusting unit cell representations and symmetry constraints, this framework could be readily extended to other lattice configurations (e.g., honeycomb and triangular lattices) using pixel-based geometric representations. Although the physical mechanism governing topological states differ across lattice types, the same inverse design framework can be applied to achieve both BGs and topological state frequencies. This underscores the versatility and robust generalizability of the proposed framework for designing topological PnCs. Beyond these lattice configurations, the framework can be readily extended to the design and optimization of three-dimensional topological PnCs and topological elastic and electromagnetic systems, paving the way for the intelligent design of complex topological materials.





\section*{Acknowledgments}

This work is supported by National Natural Science Foundation of China (Grant Nos. 12372087, U25A20189) and China Scholarship Council (Grant N0. 20250790120). V.R-G acknowledges the funding from grant CORVIBSOUND PID2023-146237NB-I00 funded by MICIU/AEI/ 10.13039/501100011033.
V.F.D.P and V.R.G are funded by the European Union’s Horizon programme in the framework of the MSCA ``META-SONIC'' project under grant agreement No. 101202648.

\section*{Data availability}

The data that support the findings of this study are available from the corresponding author upon reasonable request.

%

%
\appendix
\section{Inverse design results for bulk BGs} \label{App1}
\begin{figure*}[htbp]
    \centering
    \includegraphics[scale=1]{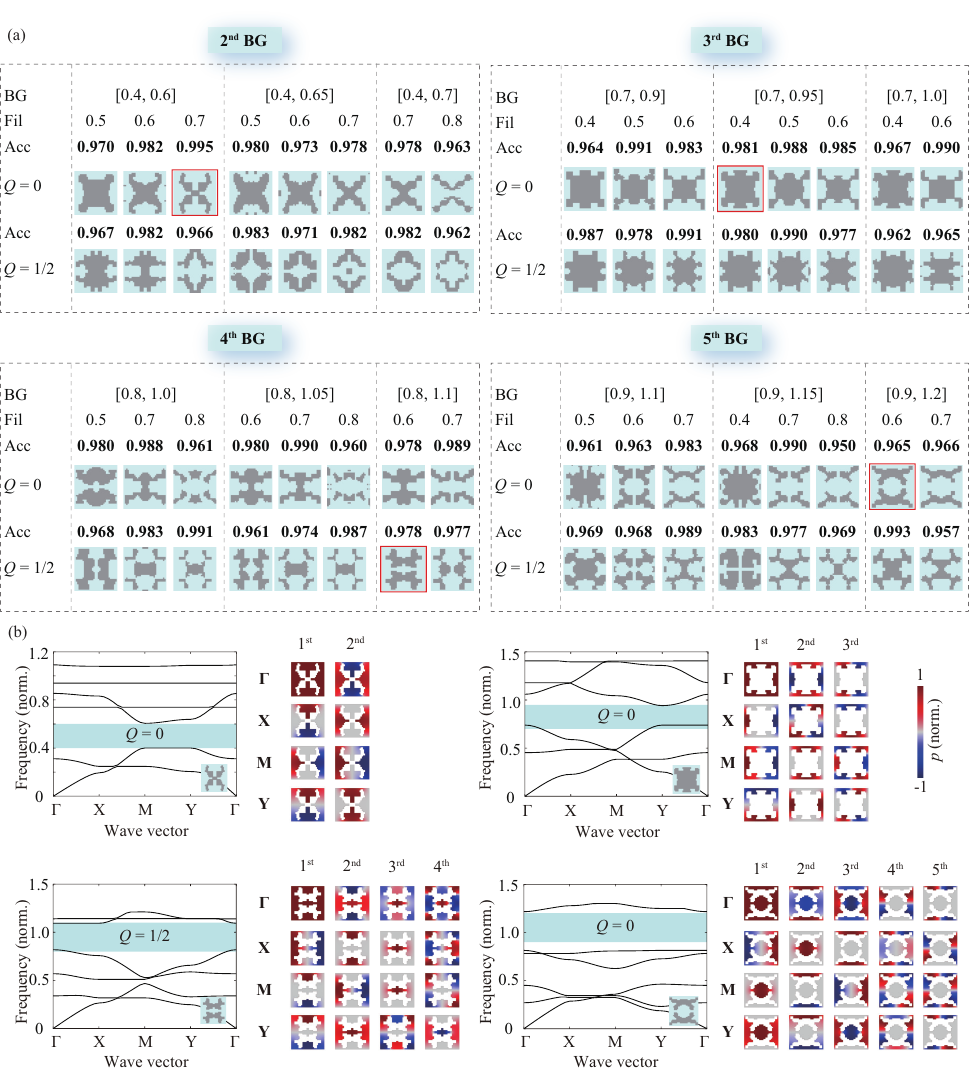}
    \captionof{figure}{\label{Fig.Inver_BG2345_Samples}(a) Representative generated unitcells for the inverse design of the $2^{\mathrm{nd}}$ to $5^{\mathrm{th}}$ bulk BG under different target BG boundaries, filling ratios, and topological indices, together with the corresponding design accuracies. (b) Band structures and eigenmodes of the representative unitcells highlighted in (a). The blue shaded regions indicate the target BG ranges, and the calculated topological indices agree with the design targets.}
\end{figure*}

For the $2^{\mathrm{nd}}$ to $5^{\mathrm{th}}$ bulk BGs, different design targets are further specified for each BG order, see Fig. \ref{Fig.Inver_BG2345_Samples}(a). In addition, for each BG order, one representative generated unitcell image is selected to examine its band structure, as shown in Fig. \ref{Fig.Inver_BG2345_Samples}(b). According to the parity of the eigenmodes below the BG at the high-symmetry points, the topological index of the designed BG can also be identified as consistent with the target value. These results demonstrate that the proposed method can achieve accurate inverse design for different bulk BGs.

\section{Details of deep learning Models} \label{App2}
\begin{figure*}[htbp]
    \centering
    \includegraphics[scale=1]{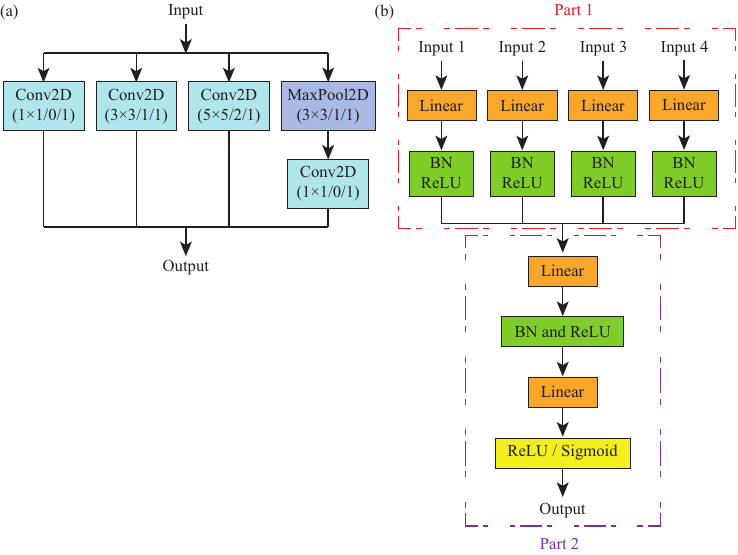}
    \caption{\label{Fig.Detail_NN}(a) Details of the Inception block. (b) Schematic illustration of the forward and generation networks.}
\end{figure*}

The detailed architecture of the conditional variational autoencoder is summarized in Table \ref{tab:CVAE}, while the structure of the Inception block employed in the encoder is illustrated in Fig. \ref{Fig.Detail_NN}(a). The Inception block extracts multiscale features through parallel convolutional branches with different kernel sizes, thereby enhancing the feature representation capability of the network. Figure \ref{Fig.Detail_NN}(b) illustrates the general architectures of the forward and generation networks used in this study. Both networks consist of two functional parts. Part 1 extracts high-dimensional features from different input variables through parallel branches, while Part 2 fuses the extracted features and performs nonlinear mapping to predict or generate the target outputs. Depending on the specific task, the number of input branches in Part 1 may vary. For some tasks, Part 1 is omitted and the input variables are directly concatenated and fed into Part 2. The detailed network configurations for different target properties are provided in Tables \ref{tab:forward} and \ref{tab:generation}.

\begin{table*}
    \centering
    \caption{Architecture of the conditional variational autoencoder}\label{tab:CVAE}
    \begin{tabular}{cccc}
        \hline
        Type & Kernel or pool size/pad/stride & Channel & Output size \\
        \hline
        Conv2D & 3$\times$3/1/1 & 32 & 16$\times$16$\times$32 \\
        MaxPool2D & 2$\times$2/0/1 & 32 & 15$\times$15$\times$32 \\
        Inception block & -- & 8-16-16-8 & 15$\times$15$\times$48 \\
        Conv2D & 3$\times$3/1/1 & 48 & 15$\times$15$\times$48 \\
        MaxPool2D & 2$\times$2/0/2 & 48 & 7$\times$7$\times$48 \\
        Inception block & -- & 16-32-32-16 & 7$\times$7$\times$96 \\
        Conv2D & 3$\times$3/1/1 & 96 & 7$\times$7$\times$96 \\
        MaxPool2D & 2$\times$2/0/2 & 96 & 3$\times$3$\times$48 \\
        Flatten & -- & -- & 1$\times$1$\times$864 \\
        Conditional branch & -- & 128-512 & 1$\times$1$\times$512 \\
        Concatenation & -- & -- & 1$\times$1$\times$1376 \\
        Fully connected & -- & 1024-512-128 & 1$\times$1$\times$128 \\
        
        Linear $\times$ 2 ($\alpha$ and $\sigma$) & -- & 10 & 1 $\times$ 1 $\times$ 10 $\times$ 2 \\
        Draw latent vector ($z$) & -- & -- & 1 $\times$ 1 $\times$ 10 \\
        
        Fully connected & -- & 256-512-1024 & 1$\times$1$\times$1024 \\
        Conditional branch & -- & 128-512 & 1$\times$1$\times$512 \\
        Concatenation & -- & -- & 1$\times$1$\times$1536 \\
        Deonv2D & 4$\times$4/0/2 & 512 & 4$\times$4$\times$512 \\
        Deonv2D & 2$\times$2/0/2 & 256 & 8$\times$8$\times$256 \\
        Deonv2D & 2$\times$2/0/2 & 1 & 16$\times$16$\times$1\\
        Sigmoid &-- & 1 & 16$\times$16$\times$1\\
        \hline
    \end{tabular}
\end{table*}

\begin{table}[!htbp]
    \centering
    \caption{Details of the forward network}\label{tab:forward}
    \begin{tabular}{ccccc}
        \hline
        Target & Input & Output & Part 1 & Part 2 \\
        \hline
        BG boundary & 10+1+1 & 10 & 512/512/512 & 512-256-128 \\
        $Q$ ($1^{\mathrm{st}}$) & 12 & 1 & -- & 32-32 \\
        $Q$ ($2^{\mathrm{nd}}$) & 12 & 1 & -- & 32-32 \\
        $Q$ ($3^{\mathrm{rd}}$) & 10+1+1 & 1 & 128/128/128 & 128-64 \\
        $Q$ ($4^{\mathrm{th}}$) & 10+1+1 & 1 & 128/128/128 & 256-256-64 \\
        $Q$ ($5^{\mathrm{th}}$) & 10+1+1 & 1 & 256/256/256 & 512-256-128-64 \\
        TIS & 20+2+2 & 2 & 256/64/64-512/512/512 & 512-256-128 \\
        TCS & 20+2+2 & 1 & 256/64/64-512/512/512 & 512-256-128 \\
        Filling ratio & 10 & 1 & -- & 512-512-256-128 \\
        \hline
    \end{tabular}
\end{table}

\begin{table}[!htbp]
    \centering
    \caption{Architecture of the generation network}\label{tab:generation}
    \begin{tabular}{ccccc}
        \hline
        Target & Input & Output & Part 1 & Part 2 \\
        \hline
        $1^{\mathrm{st}}$ BG & 1+2+1+1 & 10 & 64/64/64/64 & 256-64-32 \\
        $2^{\mathrm{nd}}$ BG & 1+2+1+1 & 10 & 64/64/64/64 & 256-64-32\\
        $3^{\mathrm{rd}}$ BG & 1+2+1+1 & 10 & 128/128/128/128 & 512-256-64-32 \\
        $4^{\mathrm{th}}$ BG & 1+2+1+1 & 10 & 128/128/128/128 & 512-256-64-32 \\
        $5^{\mathrm{th}}$ BG & 1+2+1+1 & 10 & 128/128/128/128 & 512-256-64-32 \\
        TIS & 2+2+2 & 20 & 128/128/128 & 512-256-64-32 \\
        TCS & 1+2+2 & 20 & 128/128/128 & 512-256-64-32 \\
        \hline
    \end{tabular}
\end{table}

\clearpage

\providecommand{\noopsort}[1]{}\providecommand{\singleletter}[1]{#1}%

\end{document}